\documentclass[aoas,nameyear,dvips]{arximspdf}
\usepackage{algorithm,dcolumn}
\usepackage{graphicx}

\doi{10.1214/09-AOAS314}
\volume{4}
\issue{2}
\pubyear{2010}
\firstpage{764}
\lastpage{790}

\makeatletter
\newtheorem{theorem}{Theorem}
\newtheorem{proposition}{Proposition}

\def\sfrac#1#2{#1/#2}
\newcommand{\X}{\mathbf{X}}
\newcommand{\tX}{\mathbf{X}^{T}}
\newcommand{\Delt}{\bolds{\Delta}}
\newcommand{\Delti}{\bolds{\Delta}^{-1}}
\newcommand{\U}{\mathbf{U}}
\newcommand{\Sig}{\bolds{\Sigma}}
\newcommand{\Sigi}{\bolds{\Sigma}^{-1}}
\newcommand{\Omeg}{\bolds{\Omega}}
\newcommand{\D}{\mathbf{D}}
\newcommand{\C}{\mathbf{C}}
\newcommand{\E}{\mathrm{E}}
\newcommand{\Cov}{\operatorname{Cov}}
\newcommand{\tU}{\mathbf{U}^{T}}
\newcommand{\V}{\mathbf{V}}
\newcommand{\tV}{\mathbf{V}^{T}}
\newcommand{\M}{\mathbf{M}}
\newcommand{\tr}{\operatorname{tr}}

\def\btitle#1{#1}
\def\bnote#1{#1}
\def\bpublisher#1{#1}
\def\btype#1{#1}
\def\binstitution#1{#1}
\def\bseries#1{#1}
\def\eqref#1{(\ref{#1})}
\newcolumntype{d}[1]{D{.}{.}{#1}}
\makeatother

\begin{document}
\begin{frontmatter}

\title{Transposable regularized covariance models with an application to missing data imputation}
\runtitle{Transposable regularized covariance models}

\begin{aug}
\author[A]{\fnms{Genevera I.} \snm{Allen}\corref{}\ead[label=e1]{giallen@stanford.edu}} and
\author[A]{\fnms{Robert} \snm{Tibshirani}\ead[label=e2]{tibs@stanford.edu}}
\runauthor{G. I. Allen and R. Tibshirani}
\affiliation{Stanford University}
\address[A]{Department of Statistics \\
Stanford University \\
Stanford, California 94305 \\
USA \\
\printead{e1}
\\
\phantom{E-mail: }\printead*{e2}} %adresu isvedimo komanda gale!
%Stanford University \\
%Stanford, California, 94305 \\
%USA \\
\end{aug}

% HISTORY:
\received{\smonth{4} \syear{2009}}
\revised{\smonth{11} \syear{2009}}

% ABSTRACT
\begin{abstract}
Missing data estimation is an important challenge with
high-dimensional data arranged in the form of a matrix.  Typically
this data matrix is \textit{transposable},
meaning that either the rows, columns or both can be treated as features.
To model transposable data, we
present a modification of the matrix-variate normal, the
\textit{mean-restricted matrix-variate normal},
in which the rows and columns each have a separate mean vector and
covariance matrix.
By placing additive penalties on the inverse covariance matrices
of the rows and columns, these so-called transposable regularized
covariance models allow
for maximum likelihood estimation of the mean and nonsingular
covariance matrices.
% We extend regularized covariance models, which place
% an additive penalty on the inverse covariance matrix, to this
% distribution, by placing separate penalties on the covariances of the
% rows and columns.  These so called transposable regularized covariance models allow
% for maximum likelihood estimation of the mean and nonsingular
% covariance matrices.
Using these models, we formulate EM-type
algorithms for missing data imputation
in both the multivariate and transposable frameworks.
We present theoretical results exploiting the structure of our
transposable models that allow these models and imputation methods
to be applied to high-dimensional data.
% Exploiting the structure of our transposable models, we present
% theretical results that give computational techniques enabling the
% application of our model to high-dimensional data.
% techniques enabling use of our models with high-dimensional data and
% give a
% computationally feasible one-step approximation
% for imputation.
Simulations and results on microarray data and the Netflix data show that these imputation
techniques often outperform existing methods and offer a greater degree of
flexibility.
\end{abstract}

% KEYWORDS

\end{frontmatter}

%s1 ###
\section{Introduction}\label{section_intro}

As large data sets have become more common in
biological and data mining applications, missing data imputation is a
significant challenge.  We motivate missing data estimation in matrix
data with the
example of the Netflix movie rating data [\citet{netflix}].  This data set
has around 18,000 movies (columns) and several hundred thousand
customers (rows). Customers have rated some of the movies,
but the  data  matrix is very sparse with a only small percentage
of the ratings present.
The goal is to predict the ratings for unrated movies
so as to better recommend movies to customers.  The movies and
customers, however, are very correlated and an imputation method
should take advantage of these relationships.  Customers who enjoy
horror films, for example, are likely to rate movies similarly, in the
same way that horror films are likely to have similar ratings from
these customers.
Modeling the ratings by the relationships between only the movies or
only the customers, as with multivariate methods and $k$-nearest
neighbor methods, seems shortsighted.  Customer A's rating of Movie 1,
for example, is related to Customer B's rating of Movie 2 by more than
simply the connection between Customer A and B or Movie 1 and 2.  In
addition, modeling ratings as a linear combination of the ratings of
movies or a combination of customer ratings as with singular value
decomposition (SVD) methods fails to capture a more sophisticated
connection between the movies and customers [\citet{svdknn}].
Bell et al., in their discussion of imputation for the Netflix
data, call all of these methods either
``movie-centric'' or ``user-centric'' [\citet{bellkor}].

We propose to directly model the correlations among and between both
the customers (rows) and the movies (columns). Thus, our
model is \textit{transposable} in the sense that it treats both the rows
and columns as features of interest.
The model is based on the
matrix-variate
normal distribution brought to our attention by \citet{brad1},
which has separate covariance matrix parameters
for both the
rows and the columns.  Thus, both the relationships between customers
and between movies are incorporated in the model.
If matrix-variate normal data is strung out in a long
vector, then it is distributed as multivariate normal with the covariance
related to the original row and column
covariance matrices through their Kronecker product.  This means that the
relationship between Customer A's rating of Movie~1 and Customer B's
rating of Movie 2 can be modeled directly as the interaction between
Customers A and B and Movies 1 and 2.

In practice,
however, transposable models based on the matrix-variate normal
distribution have largely been of theoretical interest and have
rarely applied to real data sets because of the computational burden of
high-dimensional parameters [\citet{gupta}].
In this paper we introduce modifications of the matrix-variate normal
distribution, specifically restrictions on the means and penalties on
the inverse covariances, that allow us to fit this transposable model to a single
matrix of data.  The penalties we employ give us nonsingular
covariance estimates that have
connections to the singular value decomposition and graphical models.
With this
theoretical foundation, we present computationally efficient
Expectation Maximization-type
(EM) algorithms for missing data imputation.  We also develop
a two-step process for calculating conditional distributions and an
algorithm for calculating conditional expectations of scattered
missing data that has the
computational cost of comparable multivariate methods.  These
contributions allow one to fit this parametric transposable model to a
single data matrix at reasonable computational cost, opening the door
to numerous applications including user-ratings data.

We organize the paper beginning with a review of the multivariate regularized
covariance models (RCM) and a new imputation method based on these models,
Section~\ref{section_RCM}.   The RCMs form the foundation
for the transposable regularized covariance models (TRCM) introduced in Section
\ref{section_TRCM}.  We then present new EM-type imputation algorithms
for transposable data,
Section~\ref{section_imp_trans}, along with a one-step approximation in Section~\ref{section_approx}.
Simulations and results on microarray and the Netflix
data are given in Section~\ref{section_res_sims}, and we conclude with a
discussion of our methods in Section~\ref{section_discussion}.

%s2 ###
\section{Regularized covariance models and imputation with multivariate data}\label{section_RCM}

Several recent papers have presented algorithms and discussed
applications of regularized covariance models (RCM) for the multivariate
normal distribution [\citet{glasso}; \citet{scout}]. These models
regularize the maximum likelihood estimate of the covariance matrix by
placing an additive penalty on the inverse covariance or concentration
matrix.  The resulting estimates are nonsingular, thus
enabling covariance estimation when the number of features is greater
than the number of observations.  In this section we give a review
of these models and briefly describe a new penalized EM algorithm for
imputation of missing values using the regularized covariance model.

Let $X_{i} \sim N(0, \Delt)$ for $i = 1,\ldots,n$, i.i.d. observations
and $p$ features.  Thus, our data matrix, $\X$, is $n \times p$ with
covariance matrix $\Delt \in \Re^{p \times p}$.  The penalized
log-likelihood of the regularized covariance model is then proportional to
\begin{equation}\label{pen_mvn_ll}
\ell( \Delt ) = \frac{n}{2} \log| \Delti |  - \frac{1}{2} \operatorname{tr} ( \tX \X\Delti )  - \rho \| \Delti \|^{q},
\end{equation}
where $\| \cdot \|^{q} = \sum_{i=1}^{p^{2}} | \cdot |^{q}$ and $q$ is
either 1 or 2, that is, the sum of the absolute value or square of the
elements of $\Delti$.  The penalty parameter is $\rho$.\vspace*{1pt}  With an
$L_{2}$ penalty, we can write the penalty term as $\rho  \operatorname{tr}(
\Delti \Delti ) = \rho \| \Delti \|_{\mathrm{F}}^{2}$.

Maximizing $\ell ( \Delt)$ gives the penalized-maximum likelihood
estimate (MLE) of $\Delt$.  \citet{glasso} present the graphical lasso algorithm to solve the problem with
an $L_{1}$ penalty.  The graphical lasso uses the lasso method
iteratively on the rows of $\hat{\Delt}{}^{-1}$, and gives a sparse
solution for $\hat{\Delt}{}^{-1}$.  A~zero in the $ij$th component of
$\Delti$ implies that variables $i$ and $j$ are conditionally
independent given the other variables.  Thus, these penalized-maximum
likelihood models with $L_{1}$ penalties can be used to estimate sparse undirected graphs.
With an $L_{2}$ penalty, the problem has an
analytical solution [\citet{scout}].  If we take the singular value
decomposition (SVD) of $\X$, $\X = \U \mathbf{D}
\tV$, with $d = \operatorname{diag}( \mathbf{D} )$, then
\begin{equation}\label{mvn_l2}
\hat{\Delt} = \V \operatorname{diag}( \theta ) \tV,\qquad
\theta_{i} = \frac{d_{i}^{2} + \sqrt{d_{i}^{4} + 16n \rho}}{2n}.
\end{equation}
Thus, the inclusion of the $L_{2}$ penalty simply regularizes the
eigenvalues of the covariance matrix.  When $p > n$ and letting $r$ be
the rank of $\X$, the final $n - r$ values of $\theta$
are constant and are equal to $2 \sqrt{\rho / n}$.  While a rank-$k$ SVD
approximation uses only the first $k$ eigenvalues, the $L_{2}$ RCM
gives a covariance estimate with all nonzero eigenvalues.
Regularized covariance models provide an alternative method of
estimating the covariance matrix with many desirable properties [\citet{spice}].

With this underlying model, we can form a new missing data imputation
algorithm by maximizing the observed penalized log-likelihood of the
regularized covariance model via the EM algorithm.  Our method is the
same as that of the EM algorithm for the multivariate normal described
in Little and Rubin [\citet{littlerubin}], except for an addition in the
maximization step.  In our M step, we find the MLE of the RCM covariance
matrix instead of the multivariate normal MLE.  Thus, our method fits
into a class of penalized EM algorithms which give nonsingular
covariance estimates [\citet{empenal}], thus
enabling use of the EM framework when $p > n$.  We give full details
of the algorithm, which we call \textit{RCMimpute}, in the~\hyperref[suppA]{Supplementary Materials} [\citet{AllenTib2010}].
As we will discuss later, this imputation
algorithm is a special case of our algorithm for transposable data and
forms an integral part of our one-step approximation algorithm
presented in Section~\ref{section_approx}.

%s3 ###
\section{Transposable regularized covariance models}\label{section_TRCM}

As previously mentioned, we model the possible dependencies between
and within the rows and columns using the matrix-variate normal
distribution.
In this section we first present a modification of this model,
the mean-restricted matrix-variate normal distribution.  We confine
the means to limit the total number of parameters and to
provide interpretable marginal distributions.  We then introduce our
transposable regularized covariance models by applying penalties to
the covariances of our
matrix-variate distribution.  Finally, we present the penalized-maximum likelihood
parameter estimates and illustrate the connections between these
estimates and those of multivariate models, the singular value decomposition
and graphical models.

%s3.1 ###
\subsection{Mean-restricted matrix-variate normal distribution}\label{section_MN}

We introduce the mean-restricted matrix-variate
normal, a variation on the  matrix-variate normal,
presented by Gupta and Nagar [\citet{gupta}].
A restriction on the means
is needed because the matrix-variate normal has a mean matrix, $\M$,
of the same dimension as $\X$, meaning that there are $n \times p$
mean parameters.  Since the matrix-variate normal is mostly applied in
instances where there are several independent samples of the random
matrix $\X$ [\citet{mnmle}], this parameter formulation is appropriate.
We propose, however, to use the model when we only have one
matrix $\X$ from which to estimate the parameters.  Also, we wish to
parameterize our model so that the marginals are multivariate normal,
thus easing computations and improving interpretability.

We denote the mean-restricted matrix-variate normal distribution by
$\X \sim N_{n,p}  ( \bolds{\nu}, \bolds{\mu}, \Sig, \Delt  )$
with $\X \in \Re^{n \times p}$, the row mean $\nu \in \Re^{n}$,
the column mean $\mu \in \Re^{p}$, the row covariance $\Sig \in
\Re^{n \times n}$ and the column covariance $\Delt \in \Re^{p
\times p}$.
If we place the matrix $\X$ into a vector of length $np$, we have
$\operatorname{vec} (\X) \sim N  ( \operatorname{vec}(\M), \bolds{\Omega}
)$,  where  $\M = \nu \mathbf{1}_{(p)}^{T} +
\mathbf{1}_{(n)} \mu^{T}$, and
$\bolds{\Omega} = \Delt \otimes \Sig$.
Thus, our mean-restricted matrix-variate normal model is a multivariate normal with a
mean matrix composed of additive elements from the row and column mean
vectors and a covariance matrix given by the Kronecker product between
the row and column covariance matrices.  This covariance structure can
be seen as a tensor product Gaussian process on the rows and
columns, an approach explored in \citet{gpnips} and \citet{srlnips}.

This distribution implies that a single element, $X_{ij}$,
has mean $\nu_{i} + \mu_{j}$ along with variance
$\Sig_{ii} \Delt_{jj}$, a mean and variance component
from the row and column to which it belongs.  As pointed out by a
referee, this can be viewed
as the following random effects model: $X_{ij} = \nu_{i} + \mu_{j} +
\varepsilon_{ij}$, where $\varepsilon_{ij} \sim N(0, \Sig_{ii} \Delt_{jj})$, which has
two additive fixed effects depending on the row and column means and a
random effect whose variance depends on the product of the
corresponding row and column covariances.  This model shares the same
first and second moments as elements from the mean-restricted
matrix-variate normal.  It does not, however, capture the Kronecker
covariance structure between the elements of $\X$ unless both the row
and column covariances, $\Sig$ and $\Delt$, are diagonal.
This random effect model differs from the more common
two-way random effects model with additive errors, which assumes that
errors from the two sources are independent.  Our model, however,
assumes that the errors are
related and models them as an interaction effect.
A similar random effects approach was
taken in \citet{reicml}, also
using a Kronecker product covariance matrix.

To further illustrate the
model, we note that the rows and columns are both marginally
multivariate normal. The $i$th row, denoted as $X_{i r}$, is
distributed as
$X_{i r} \sim N  ( \nu_{i} + \mu,
\Sigma_{ii} \Delt  )$  and the $j$th column, denoted by
$X_{c j}$, is distributed as $X_{c j} \sim N
( \nu +   \mu_{j}, \Delta_{jj} \Sig  ).$
The familiar multivariate normal distribution is a special case of the
mean-restricted matrix-variate normal as seen by the following two
statements. If $\Sig = \mathbf{I}$ and $\nu = \mathbf{0,}$  then
$\X \sim N  ( \mu, \Delt  )$, and if $\Delt = \mathbf{I}$
and $\mu = \mathbf{0,}$ then $\X \sim N  ( \nu, \Sig  )$.
Also, two
elements from different rows or columns are distributed as a bivariate
normal, $( X_{ij}, X_{i'j'} ) \sim N  (  {
\nu_{i} + \mu_{j} \choose \nu_{i'} + \mu_{j'} } ,
\bigl(
{\Sig_{ii} \Delt_{jj} \atop \Sig_{i'i} \Delt_{j'j}}
\enskip
{\Sig_{ii'} \Delt_{jj'} \atop \Sig_{i'i'}\Delt_{j'j'}}
\bigr)   )
%{\Sig_{ii} \Delt_{jj} \choose \Sig_{i'i} \Delt_{j'j}}
$.
Thus, our model is more general than the multivariate normal,
with the flexibility to encompass many different marginal
multivariate models.

For completeness, the density function of this distribution is
\begin{eqnarray*}
&&p  ( \nu, \mu, \Sig, \Delt  )\nonumber
\\
&&\qquad= (2 \pi)^{-\sfrac{np}{2}}   | \Sig |^{-\sfrac{p}{2}}    | \Delt|^{-\sfrac{n}{2}} \\
&&\qquad\quad{}\times \operatorname{etr}    \bigl( - \tfrac{1}{2} \bigl(\X - \nu \mathbf{1}_{(p)}^{T} - \mathbf{1}_{(n)} \mu^{T}\bigr) \Delti
\bigl(\X- \nu \mathbf{1}_{(p)}^{T} -\mathbf{1}_{(n)} \mu^{T}\bigr)^{T} \Sigi  \bigr),\nonumber
\end{eqnarray*}
where  $\operatorname{etr}(\cdot)$ is the exponential of the trace function.
Hence, our formulation of the matrix-variate normal distribution adds
restrictions on the means, giving the distribution desirable properties in
terms of its marginals and easing computation of parameter estimates,
discussed in the following section.

%s3.2 ###
\subsection{Transposable Regularized Covariance Model (TRCM)}\label{section_sub_TRCM}

In the previous section we have reformulated the distribution to
limit the mean parameters
and in this section we regularize the covariance parameters.
This allows us to obtain nonsingular covariance estimates which are
important for use in any application, including missing data
imputation.

As in the multivariate case, we seek to penalize the inverse
covariance matrix.  Instead of penalizing the overall covariance,
$\Omeg$, we add two separate penalty terms, penalizing the inverse
covariance of the rows and of the columns.  The
penalized log-likelihood is thus
\begin{eqnarray}\label{pen_mn_ll}
&&\ell(\nu, \mu, \Sig, \Delt)\nonumber
\\
&&\qquad= \frac{p}{2} \log |\Sigi|   +\frac{n}{2} \log |\Delti| \nonumber
\\[-8pt]\\[-8pt]
&&\qquad\quad{}    -  \frac{1}{2} \operatorname{tr} \bigl( \Sigi \bigl(\X  - \nu \mathbf{1}_{(p)}^{T} -
\mathbf{1}_{(n)} \mu^{T}\bigr)  \Delti\bigl(\X - \nu \mathbf{1}_{(p)}^{T} -
\mathbf{1}_{(n)} \mu^{T}\bigr)^{T}  \bigr) \nonumber \\
&&\qquad\quad{}    - \rho_{r}\|\Sigi\|^{q_{r}} - \rho_{c}\|\Delti\|^{q_{c}},\nonumber
\end{eqnarray}
where $\| \cdot \|^{q_{r}} = \sum_{i=1}^{m^{2}} | \cdot |^{q_{r}}$ and
$q_{r}$ and $q_{c}$ are either 1 or 2, that is, the sum of the absolute
value of the matrix elements or squared elements.  $\rho_{r}$ and
$\rho_{c}$ are the two penalty parameters.
Note that we will refer to the four possible types of penalties as
$L_{q_{r}}\dvtx L_{q_{c}}$.
Placing separate penalties on the two covariance matrices is
not equivalent to placing a single penalty on the Kronecker product
covariance matrix $\bolds{\Omega}$.  Using two separate penalties
gives greater flexibility, as the covariance of the rows and columns
can be modeled separately using differing penalties and penalty parameters.  Also, having two
penalty terms leads to simple parameter estimation strategies.

With transposable regularized covariance models, as with their
multivariate counterpart, the penalties are placed on the inverse
covariance matrix, or concentration matrix.  Estimation of the
concentration matrix has long been associated with graphical
models, especially with an $L_{1}$ penalty which is useful to model
sparse graphical models [\citet{glasso}].
Here, a nonzero entry of the concentration matrix, $\Sig_{ij} \neq
0$, means that the $i$th row conditional on all other rows is
correlated with row $j$.  Thus, a ``link'' is formed in the graph
structure between nodes $i$ and $j$.  Conversely, zeros in the
concentration matrix imply conditional independence.  Hence, since we
are estimating both a regularized row and column concentration matrix,
our model can be interpreted as modeling both the rows and columns
with a graphical model.

%s3.3 ###
\subsection{Parameter estimation}\label{section_param_est}

We estimate the means and covariances via penalized maximum likelihood
estimation. The estimates, however, are not unique, but the overall mean,
$\hat{\M}$, and overall covariance $\hat{\Omeg}$ are unique.  Hence, $\hat{\nu}$ and
$\hat{\mu}$ are unique up to an additive constant and $\hat{\Sig}$ and $\hat{\Delt}$ are
unique up to a multiplicative constant.  We first begin with the
maximum likelihood estimation of the mean parameters.
%p1
\begin{proposition}\label{mean_mn}
The MLE estimates for $\nu$ and $\mu$ are
\begin{eqnarray}\label{mn_mean}
\hat{\nu} = \sum_{j=1}^{p} \frac{(X_{c j} - \hat{\mu}_{j})}{p},
\qquad
\hat{\mu} = \sum_{i=1}^{n} \frac{(X_{i r} - \hat{\nu}_{i})}{n},
\end{eqnarray}
where $X_{c j}$ denotes the $j$th column and $X_{i
r}$ the $i$th row of $\X \in \Re^{n \times p}$.
\end{proposition}
\begin{pf}
See~\hyperref[suppA]{Supplementary Materials}.
\end{pf}

The estimates for $\nu$ and $\mu$ are obtained by centering with
respect to the rows and then the columns.  Note that centering by the
columns first will change $\hat{\mu}$ and $\hat{\nu}$, but will still
give the same additive result.  Thus, the order in which we center is
unimportant.

Maximum likelihood estimation of the covariance matrices is more
difficult.  Here, we will assume that the data has been centered, $\M
= \mathbf{0}$.  Then, the penalized log-likelihood, $\ell( \Sig, \Delt )$, is a
bi-concave function of $\Sigi$ and $\Delti$.  In words, this means that
for any fixed ${\bolds{\Sigma}^{-1}}{}'$, $\ell(\bolds{\Sigma}', \Delt)$ is a concave function of
$\Delti$, and for any fixed ${\bolds{\Delta}^{-1}}{}'$, $\ell(\Sig, \bolds{\Delta}')$ is a
concave function of $\Sigi$.  We exploit this structure to maximize
the penalized likelihood by iteratively maximizing along each
coordinate, either $\Sigi$ or $\Delti$.
%p2
\begin{proposition}\label{cw_conv}
Iterative block coordinate-wise maximization of $\ell( \Sig, \Delt)$
with respect to $\Sigi$ and $\Delti$ converges to a stationary point
of $\ell(\Sig, \Delt)$ for both $L_{1}$ and $L_{2}$ penalty types.
\end{proposition}
\begin{pf}
See~\hyperref[suppA]{Supplementary Materials}.
\end{pf}

While block coordinate-wise maximization (Proposition~\ref{cw_conv})
reaches a stationary point of $\ell(\Sig, \Delt)$, it is not
guaranteed to reach the global maximum.  There are potentially many
stationary points, especially with $L_{1}$ penalties, due to the
high-dimensional nature of the parameter space.  We also note a few
straightforward
properties of the coordinate-wise maximization procedure, namely, that
each iteration monotonically increases the penalized log-likelihood
and the order of maximization is unimportant.

The coordinate-wise maximization is accomplished by setting the
gradients with respect to $\Sigi$ and $\Delti$ equal to zero and
solving.  We list the gradients with $L_{2}$ penalties.  With $L_{1}$
penalties, only the third term is changed and is given in parentheses:
\begin{eqnarray}\label{l2_grads}
\frac{\partial \ell}{\partial \Sigi} &=& \Sig - \X \Delti \tX /p -
\frac{4 \rho_{r}}{p} \Sigi   \biggl( \frac{2
\rho_{r}}{p} \operatorname{sign}(\Sigi) \biggr), \nonumber \\[-8pt]\\[-8pt]
\frac{\partial \ell}{\partial \Delti} &=& \Delt - \tX \Sigi \X /n -
\frac{4 \rho_{c}}{n}  \Delti  \biggl( \frac{2 \rho_{c}}{n}
\operatorname{sign}(\Delti)  \biggr).\nonumber
\end{eqnarray}
Maximization with $L_{1}$ penalties can be achieved by applying the
graphical lasso
algorithm to the second term with the coefficient of the third term
as the penalty parameter.
With $L_{2}$ penalties, we maximize by taking the eigenvalue decomposition of
the second term and regularizing the eigenvalues as in the
multivariate case, \eqref{mvn_l2}.  Thus, coordinate-wise maximization
leads to a simple iterative algorithm, but it comes at a cost since it
does not necessarily converge to the global maximum.  When both
penalty terms are $L_{2}$ penalties, however, we can find the global maximum.

%s3.3.1 ###
\subsubsection{Covariance estimation for $L_{2}$ penalties}\label{section_cov_est_l2}

Covariance estimation\break when both penalties of the
transposable regularized covariance model are $L_{2}$ penalties
reduces to a minimization problem involving the eigenvalues of the
covariance matrices.  This problem has a unique analytical solution and, thus,
our estimates, $\hat{\Sig}$ and $\hat{\Delt}$, are globally
optimal.
%t1
\begin{theorem}\label{l2l2_sol}
The global unique solution maximizing $\ell(\Sig, \Delt)$ with $L_{2}$
penalties on both covariance parameters is given by the following: Denote the SVD of $\X$ as $\X = \U
\D \tV$  with $d = \operatorname{diag}(\D)$ and let $r$ be the rank of $\X$, then
\begin{eqnarray}\label{l2_sol}
\Sig^{*} = \U \operatorname{diag}( \beta^{*} ) \tU  \quad\mbox{and}\quad
\Delt^{*} = \V \operatorname{diag}( \theta^{*} ) \tV,
\end{eqnarray}
where $\beta^{*} \in \Re^{n+}$ and $\theta^{*} \in \Re^{p+}$ given by
\begin{eqnarray*}
\beta^{*}_{i} &=&
\cases{
\displaystyle 2 \sqrt{ \frac{\rho_{r}}{p} }, &\quad \mbox{if }$i \geq r$,\vspace*{4pt} \cr
\displaystyle\sqrt{ \frac{- c_{2}^{(i)} - \sqrt{ {c_{2}^{(i)}}{}^{2} - 4 c_{1}^{(i)} c_{3}^{(i)}} }{2 c_{1}^{(i)}}}, &\quad \mbox{otherwise}, \nonumber
}
\\
\theta^{*}_{i} &=&
\cases{
\displaystyle2 \sqrt{\frac{\rho_{c}}{n}}, &\quad \mbox{if }$i \geq r$,\vspace*{4pt} \cr
\displaystyle\frac{ d_{i}^{2} \beta^{*}_{i}}{p {\beta^{*}_{i}}{}^{2} - 4 \rho_{r} }, &\quad \mbox{otherwise},
}
\end{eqnarray*}
with coefficients
\begin{eqnarray*}
c_{1}^{(i)} &=& -4 \rho_{c} p^{2},\qquad
c_{2}^{(i)} = 32 \rho_{r} \rho_{c} p + d_{i}^{4} ( n - p),
\quad\mbox{and}\quad
\\
c_{3}^{(i)} &=& 4 \rho_{r} ( d_{i}^{4} - 16 \rho_{r} \rho_{c} ).
\end{eqnarray*}
\end{theorem}
\begin{pf}
See~\hyperref[suppA]{Supplementary Materials}.
\end{pf}

With $L_{2}$ penalties, maximum likelihood covariance estimates $\hat{\Sig}$ and
$\hat{\Delt}$ have eigenvectors given by the left and right
singular vectors of $\X$, respectively.  To reveal some intuition as to
how these covariance estimates compare to other possible eigenvalue
regularization methods, we present the two gradient equations in terms
of the eigenvalues $\beta$ and $\theta$  (these are discussed fully in the proof of Theorem~\ref{l2l2_sol}):
\begin{eqnarray*}
p \theta_{i} \beta_{i}^{2} - d_{i}^{2} \beta_{i} - 4 \rho_{r} \theta_{i}  = 0
\quad\mbox{and}\quad
n \beta_{i} \theta_{i}^{2} - d_{i}^{2} \theta_{i} - 4 \rho_{c} \beta_{i}  = 0.
\end{eqnarray*}
These are two quadratic functions in $\beta$ and $\theta$, so
the quadratic formula gives us the eigenvalues in terms of each
other.
% {\small \begin{align*}
% \beta = \frac{d^{2} + \sqrt{d^{4} + 16 p \rho+{r} \theta }}{2 p
%   \theta}  \&
% \theta = \frac{d^{2} + \sqrt{d^{4} + 16 n \rho_{c} \beta}}{2 n \beta}.
% \end{align*}}
We see that the eigenvalues regularize the square of the
singular values by a function of the dimensions, the penalty parameters
and the eigenvalues of the other covariance estimate.
From Theorem~\ref{l2l2_sol}, $L_{2}\dvtx L_{2}$ covariance estimation has a unique and globally
optimal solution, which cannot be said of the other combinations of
penalties.  We give numerical results comparing our TRCM
covariance estimates to other shrinkage covariance estimators in the
\hyperref[suppA]{Supplementary Materials}.

Here, we also pause to compare our TRCM model with $L_{2}$ penalties
to the singular value decomposition model commonly employed with matrix
data.  If we include both row and column intercepts, we can write the rank-reduced
SVD model as $X_{ij} = \nu_{i} + \mu_{j} +
\mathbf{u}_{i}^{T}\mathbf{D}_{r} \mathbf{v}_{j} + \varepsilon$, where
$\mathbf{u}_{i}$ and $\mathbf{v}_{j}$ are the $i$th and $j$th
right and left singular vectors, $\mathbf{D}_{r}$ is the rank-reduced
diagonal matrix of singular values and $\varepsilon \sim N(0,
\sigma^{2})$.  Thus, the model appears similar to $L_{2}$ TRCM, which
can be written as $X_{ij} = \nu_{i} + \mu_{j} + \varepsilon_{ij}$ where
$\varepsilon_{ij} \sim N( 0, \mathbf{u}^{T}_{i} \operatorname{diag}(\beta)
\mathbf{u}_{i} * \mathbf{v}_{j}^{T} \operatorname{diag}(\theta)
\mathbf{v}_{j} )$. There are important differences between the models,
however.  First, the left and right singular vectors are incorporated
directly into the SVD model, whereas they form the bases of the
variance component of TRCM.  Second, a rank-reduced SVD incorporates
only the first $r$ left and right singular vectors.  Our model uses
all the singular vectors as $\beta$ and $\theta$ are of lengths $n$
and $p$, respectively.  Finally, the SVD allows the covariances of the
rows to vary with $\mathbf{u}_{i}$, whereas with TRCM the rows share a
common covariance matrix.  Thus, while the SVD and TRCM share
similarities, the models differ in structure and, hence, each offers a
separate approach to matrix-data.

%s4 ###
\section{Imputation for transposable data}\label{section_imp_trans}

Imputation methods for transposable data are the main
focus of this
paper.  We formulate methods based on the transposable
regularized covariance models introduced in Section~\ref{section_TRCM}.
Because computational costs have limited use of the
matrix-variate normal in applications, we let
computational considerations motivate
the formulation of our imputation methods.

We propose a Multi-Cycle Expectation Conditional Maximization\break
(MCECM)
algorithm, given by \citet{ecm},
maximizing the observed penalized log-likelihood of
the transposable regularized covariance models.  The algorithm exploits the structure of our model by
maximizing with respect to one block of coordinates at a time, saving
considerable mathematical and computational time.  First, we develop
the algorithm mathematically, provide some
rationale behind the structure of the algorithm via numerical
examples, and then briefly discuss computational
strategies and considerations.

In high-dimensional data, however, the MCECM algorithm we propose
for imputation is not computationally feasible.  Hence, we suggest a
computation-saving one-step approximation in
Section~\ref{section_approx}.  The foundation of our approximation lies in new methods,
given in Theorems~\ref{mn_conds}  and~\ref{e_approx}, for
calculating conditional distributions with the mean-restricted
matrix-variate normal.
We also demonstrate the utility of this one-step procedure in numerical
examples. A Bayesian
variation of the one-step approximation using Gibbs
sampling is given in the~\hyperref[suppA]{Supplementary Materials}.

Prior to formulating the imputation
algorithm for transposable models, we pause to
address a logical question:  Why do we not use the multivariate
imputation method based on regularized covariance models, given that the mean-restricted
matrix normal distribution
can be written as a multivariate normal with $\mbox{vec}(\X) \sim N( \mbox{vec}
(\M), \Omeg )$?
There are two reasons why this is inadvisable.  First, notice that
TRCMs place an additive penalty on both the inverse covariance
matrices of the rows and the columns.  The overall covariance matrix,
$\Omeg$, however, is their Kronecker product.  Thus, converting the TRCM into a
multivariate form yields a messy penalty term leading to a
difficult maximization step.  The second reason to avoid
multivariate methods is computational.  Recall that $\Omeg$ is a $np
\times np$ matrix which is expensive to repeatedly invert.  We will
see that the mathematical form of the ECM imputation algorithm we
propose leads to computational strategies that avoid the expensive
inversion of $\Omeg$.

%s4.1 ###
\subsection{Multi-cycle ECM algorithm for imputation}\label{section_MCECM}

Before presenting the algorithm, we first review the
notation used throughout the remainder of this paper.  As
previously mentioned, we use $i$ to
denote the row index and $j$ the column index.  The observed and missing parts
of row $i$ are
$o_{i}$ and $m_{i}$, respectively, and $o_{j}$ and $m_{j}$ are the
analogous parts of column $j$.  We let $m$ and $o$ denote the totality
of missing and observed elements, respectively.  Since with
transposable data there is no natural orientation, we set $n$ to
always be the larger dimension of $\X$ and $p$ the smaller.

%s4.1.1 ###
\subsubsection{Algorithm}

We develop the ECM-type algorithm for imputation mathematically,
beginning with the
observed data log-likelihood which we seek to maximize.  Letting
$x^{*}_{o_{j},j} = \Sig^{-1/2}_{o_{j},o_{j}}( x_{o_{j},j} -
\nu_{o_{j}} )$,
\begin{eqnarray}\label{mn_obs_ll}
\ell (\nu, \mu, \Sig, \Delt) &=&  \frac{1}{2}  \Biggl[ \sum_{j=1}^{p}
\log | \Sigi_{o_{j},o_{j}} | + \sum_{i=1}^{n}
|\Delti_{o_{i},o_{i}} |    \Biggr]\nonumber
\\
&&{}     - \frac{1}{2} \operatorname{tr}  \Biggl( \sum_{i=1}^{n}
(x^{*}_{i,o_{i}} - \mu_{o_{i}})^{T} (x^{*}_{i,o_{i}} -
\mu_{o_{i}}) \Delti_{o_{i},o_{i}}  \Biggr)
\\
&&{}- \rho_{r} \|\Sigi   \|^{q_{r}} - \rho_{c} \|\Delti\|^{q_{c}}.  \nonumber
\end{eqnarray}
One can show that this is indeed the observed log-likelihood by
starting with the multivariate observed log-likelihood and using $\operatorname{vec}(\X)$ and the
corresponding $\operatorname{vec}(\M)$ and $\Omeg$.  We maximize
\eqref{mn_obs_ll} via an EM-type algorithm which, similarly to the
multivariate case, gives the imputed values as a part of the Expectation step.

We present two forms of the E step, one which leads to
simple maximization with respect to $\Sigi$ and the other with respect
to $\Delti$.  This is possible because of the structure of the
matrix-variate model, specifically the trace term.  Letting $\theta =
\{ \nu, \mu, \Sig, \Delt \}$, the parameters of the mean-restricted
matrix-variate normal, and letting $o$ be the indices of the observed
values, the E step, denoted by $Q(\theta |
\theta', X_{o})$, has the following form.   Here, we
assume that $\X$ is centered:
%{\small
\begin{eqnarray*}\label{mn_q_func}
Q(\theta | \theta', X_{o} ) &=&
\mathrm{E}  ( \ell( \nu, \mu, \Sig,
\Delt) | X_{o}, \theta'  )
 \propto \E  [ \tr  ( \tX \Sigi \X \Delti  ) | X_{o},
\theta'  ] \nonumber \\
& \propto& \tr  [ \E  ( \tX \Sigi \X | X_{o}, \theta'  )
\Delti  ] \nonumber
 \propto \tr  [ \E  ( \X \Delti \tX | X_{o}, \theta'
) \Sigi  ].  \nonumber
\end{eqnarray*}
%}
Thus, we have two equivalent forms of the conditional expectation
which we give below.
%p3
\begin{proposition}\label{mn_e_step}
The E step is proportional to the following form:
%{\small
\begin{eqnarray}
\mathrm{E}  [ \operatorname{tr}  ( \tX \Sigi \X\Delti  ) | X_{o}, \theta'  ]
&=& \tr  \bigl[  \bigl( \hat{\X}^{T}\nonumber
\Sigi \hat{\X} + \mathbf{G}( \Sigi )  \bigr) \Delti  \bigr] \\[-8pt]\\[-8pt]
&=& \tr  \bigl[  \bigl( \hat{\X}
\Delti \hat{\X}^{T} + \mathbf{F}( \Delti )  \bigr) \Sigi  \bigr],
\nonumber
\end{eqnarray}
%}
%{\small
where  $\hat{\X} =  \E   ( \X | X_{o}, \theta'  )$
and
\begin{eqnarray*}
\mathbf{G}( \Sigi) &=&  \pmatrix{
\tr \bigl( \C^{(11)} \Sigi \bigr) & \cdots & \tr \bigl( \C^{(1p)} \Sigi \bigr) \cr
\vdots & \ddots & \vdots \vspace*{4pt}\cr
\tr \bigl( \C^{(p1)} \Sigi \bigr) & \cdots & \tr \bigl(\C^{(pp)} \Sigi \bigr)
},
\\
\C^{(jj')} &=& \Cov  (  X_{c j}, X_{c  j'} | X_{o}, \theta'  ), \\
\mathbf{F}( \Delti) &=&
\pmatrix{
\tr \bigl( \D^{(11)} \Delti \bigr) & \cdots & \tr \bigl( \D^{(1n)} \Delti \bigr)\cr
\vdots & \ddots & \vdots \vspace*{4pt}\cr
\tr \bigl( \D^{(n1)} \Delti \bigr) & \cdots & \tr \bigl(\D^{(nn)} \Delti \bigr) }
,
\\
\D^{(ii')} &=& \Cov  ( X_{i r}, X_{i'    r} | X_{o}, \theta'  ).
\end{eqnarray*}
\end{proposition}
\begin{pf}
See~\hyperref[suppA]{Supplementary Materials}.
\end{pf}

The E step in the matrix-variate normal framework has a similar
structure to that of the multivariate normal (see~\hyperref[suppA]{Supplementary Materials}) with an
imputation step~($\hat{\X}$) and a covariance correction step
[$\C^{(jj')}$ and $\D^{(ii')}$].
The matrices $\C^{(jj')} \in \Re^{n \times n}$ and $\D^{(ii')} \in \Re^{p
\times p}$, while $\mathbf{G}( \Sigi) \in \Re^{p \times p}$  and
$\mathbf{F}( \Delti ) \in \Re^{n \times n}$.  Note that $\C^{(jj')}$
is sparse and
only nonzero at $\C_{ii'}^{(jj')}$ when $x_{ij}$ and $x_{i'j'}$ are
both missing.   $\C^{(jj')}$ is not symmetric, but
${\C^{(jj')}}{}^{T} = \C^{(j'j)}$, hence, $\mathbf{G}(\Sigi)$ is
symmetric.  The matrices $\D^{(ii')}$ and $\mathbf{F} (\Delti )$ are
structured analogously.
Thus, we have two equivalent forms of the
E step, which will be inserted between the two Conditional Maximization
(CM) steps to form the MCECM algorithm.

The CM steps which maximize the conditional expectation functions, in
Proposition~\ref{mn_e_step}, along with either $\Sigi$ or
$\Delti$ are direct extensions of the MLE
solvers for the multivariate RCMs. This is
easily seen from the gradients.  Note that we only show the gradients with
an $L_{2}$ penalty, since an $L_{1}$ penalty differs only in the last term:
\begin{eqnarray*}
\frac{\partial Q}{\partial \Sigi} &=& \Sig -  [ \hat{\X} \Delti \hat{\X}^{T} +
\mathbf{F}(\Delti)  ]/p - \frac{4 \rho_{r} }{p}
\Sigi, \nonumber \\
\frac{\partial Q}{\partial \Delti} &=& \Delt -  [ \hat{\X}^{T} \Sigi \hat{\X} +
\mathbf{G}(\Sigi)  ]/n - \frac{4 \rho_{c} }{n} \Delti. \nonumber
\end{eqnarray*}
With an $L_{2}$ penalty,
the estimate is given by taking the eigenvalue decomposition of the
second term and regularizing the eigenvalues as in \eqref{mvn_l2}.
The graphical lasso algorithm applied to the second term gives the
estimate in the case with an $L_{1}$ penalty.
%alg1
\begin{algorithm}[b]%[!h]
\caption{Imputation with Transposable Regularized Covariance Models (\mbox{TRCMimpute})}\label{trans_imp}
\begin{enumerate}
\item Initialization:
\begin{enumerate}[(a)]
\item[(a)] Estimate $\hat{\nu}$ and $\hat{\mu}$ from the observed data.
\item[(b)] If $x_{ij}$ is missing, set $x_{ij} = \hat{\nu}_{i} + \hat{\mu}_{j}$.
\item[(c)] Start with nonsingular estimates $\hat{\Sig}$ and $\hat{\Delt}$.
\end{enumerate}
\item E Step ($\Delt$): Calculate $ \hat{\X}^{T} \hat{\Sig}{}^{-1} \hat{\X} +
\mathbf{G}(\hat{\Sig}{}^{-1}) $.
\item M Step ($\Delt$):
\begin{enumerate}[(a)]
\item[(a)] Update estimates of $\hat{\nu}$ and $\hat{\mu}$.
\item[(b)] Maximize $Q$ with respect to $\Delti$ to obtain $\hat{\Delt}$.
\end{enumerate}
\item E Step ($\Sig$): Calculate $\hat{\X} \hat{\Delt}{}^{-1} \hat{\X}^{T} +
\mathbf{F}(\hat{\Delt}{}^{-1})$,
\item M Step ($\Sig$):
\begin{enumerate}[(a)]
\item[(a)] Update estimates of $\hat{\nu}$ and $\hat{\mu}$.
\item[(b)] Maximize $Q$ with respect to $\Sigi$ to obtain $\hat{\Sig}$.
\end{enumerate}
\item Repeat Steps 2--5 until convergence.
\end{enumerate}
\end{algorithm}

We now put these steps together and present the Multi-Cycle ECM
algorithm for imputation with transposable data, \mbox{TRCMimpute}, in Algorithm
\ref{trans_imp}.  A brief comment regarding the initialization of
parameter estimates is needed.
Estimating the
mean parameters when missing values are present is not as simple as
centering the rows and columns as in \eqref{mn_mean}.
Instead, we iterate centering by rows and columns, ignoring
the missing values by summing over the observed values, until
convergence.  Second, the
initial estimates of $\hat{\Sig}{}^{-1}$ and $\hat{\Delt}{}^{-1}$ must be
nonsingular in order to preform the needed computations in the E
step.  While any nonsingular matrices will work, we find that the algorithm
converges faster if we start with the MLE estimates with the missing
values fixed and set to the estimated mean.  Some properties and
numerical comparisons of the MCECM algorithm are given in the~\hyperref[suppA]{Supplementary Materials}.

%s4.1.2 ###
\subsubsection{Computational considerations}

We have presented our imputation algorithm for
transposable data, \mbox{TRCMimpute}, but have not yet
discussed the computations required.  Calculation of the terms for
the two E steps can be especially troublesome and, thus, we concentrate
on these.
Particularly, we need to find $\hat{\X} = \E
( \X | X_{o}, \theta'  )$, and the covariance terms,
$\C^{(jj')} = \Cov  ( X_{c j}, X_{c j'} | X_{o}, \theta'
)$ and $\D^{(ii')} = \Cov  ( X_{i r}, X_{i' r} |
X_{o}, \theta'  )$.  The simplest but not always the most
efficient way to compute these is to use the multivariate normal
conditional formulas with the Kronecker covariance matrix
$\bolds{\Omega}$, that is, if we let $m$ be the indices of the missing
values of $\operatorname{vec}(\X)$ and $o$ be the observed,
\begin{eqnarray}\label{full_mn_cond}
\qquad \operatorname{vec}(\hat{\X})_{k} = \cases{
\operatorname{vec}(\M)_{k} + \bolds{\Omega}_{ko} \bolds{\Omega}^{-1}_{oo}
\bigl(\operatorname{vec}(\X)_{o} - \operatorname{vec}(\M)_{o}   \bigr),  &\quad \mbox{if }$k \in m$, \vspace*{2pt}\cr
\operatorname{vec}(X)_{k}, &\quad\mbox{if }$k \in o$,  }\vspace*{-15pt}
\end{eqnarray}
and the nonzero elements of $\C$ and $\D$ corresponding to
covariances between pairs of missing values come from
\begin{eqnarray}
\Cov  ( \operatorname{vec}(\X)_{m}, \operatorname{vec}(\X)_{m} | X_{o},
\theta'  ) =
\bolds{\Omega}_{mm} - \bolds{\Omega}_{mo} \bolds{\Omega}_{oo}^{-1}
\bolds{\Omega}_{om}.
\end{eqnarray}
This computational strategy may be appropriate for small data matrices, but
even when $n$ and $p$ are medium-sized, this approach can be
computationally expensive.  Inverting $\bolds{\Omega}$ can be of
order $O(n^{3}p^{3})$, depending on the amount of missing data.  So,
even if we have a relatively small matrix of dimension $100 \times
50$, this inversion costs around $O(10^{10})$!  Using Gibbs sampling
to approximate the calculations of the E steps in either a Stochastic
or Stochastic Approximation EM-type algorithm [\citet{sem}] is one
computational approach (we present Gibbs sampling as part of our
Bayesian one-step approximation in the~\hyperref[suppA]{Supplementary Materials}).  A
stochastic approach, however, is still computationally expensive and,
thus, an approximation to our MCECM algorithm is needed.

%s4.2 ###
\subsection{One-step approximation to \mbox{TRCMimpute}}\label{section_approx}

For high-dimensional\break transposable data, the imputation
algorithm,
\mbox{TRCMimpute}, can be computationally prohibitive.
Thus, we propose
a one-step approximation which
has computational costs comparable to multivariate imputation
methods.

%s4.2.1 ###
\subsubsection{One-step algorithm}

The MCECM algorithm for imputation with transposable regularized
covariance models iterates between the E step, taking conditional
expectations, and the CM steps, maximizing with respect to the inverse
covariances.  Both of these steps are computationally intensive for
high-dimensional data.  While each iterate increases the observed
log-likelihood, the first step usually produces the steepest increase
in the objective.  Thus, we propose an algorithm that instead of
iterating between E and CM steps, approximates the solution of the
MCECM algorithm by stopping after only one step.

Many have noted in other iterative maximum likelihood-type algorithms
that a one-step algorithm from a good initial starting point often
produces an efficient, if not comparable, approximation to the
fully-iterated solution
[\citet{cosso}; \citet{fanli}].  Thus, for our one-step approximation we seek a
good initial solution from which to start our CM and E steps.  For
this, we turn to the multivariate regularized covariance models.
Recall that all marginals of the mean-restricted matrix-variate normal
are multivariate normal and, hence, if one of the penalty parameters
for the TRCM model
is infinitely large, we obtain the RCM solution (i.e., if $\rho_{r} =
\infty$, we get the RCM solution with penalized covariance among the
columns).  We propose
to use the estimates from the two marginal distributions with
penalized row covariances and penalized column covariances to obtain our
initial starting point.  This is similar to the COSSO one-step
algorithm which uses a marginal solution as a good initial starting
point [\citet{cosso}].

Since the final goal of our approximation algorithm is missing value
imputation, and not parameter estimation, we then tailor our one-step
algorithm to
favor imputation.  First, instead of using the marginal RCM
covariance estimates as starting values for the subsequent \mbox{TRCMimpute}
E and CM steps, we use the marginal estimates to obtain two sets of
imputed missing values through applying the \mbox{RCMimpute} method to the
rows and then the columns.
We then average the two sets of missing value
estimates and fix these to find the maximum likelihood parameter
estimates for the TRCM model, completing the maximization step.
In summary, our initial estimates are obtained by applying an EM-type method
to the marginal models.  \citet{emstartvals}
similarly use other EM-type algorithms to find good initial starting
values for their EM mixture model algorithm.  The final step of our
algorithm is the Expectation step where we take the conditional
expectation of the missing values given the observed values and the
TRCM estimates.  Note that the E step of the MCECM algorithm includes
both an imputation part and a covariance correction part (see
Proposition~\ref{mn_e_step}).  For our one-step algorithm, however, the
covariance correction part is unnecessary since our final goal is
missing value imputation.
We give the one-step approximation, called
\mbox{TRCMAimpute}, in Algorithm~\ref{trans_imp_approx}.
%alg2
\begin{algorithm}%[!h]
\caption{One-step algorithm approximating \mbox{TRCMimpute} (TRCMAimpute)}\label{trans_imp_approx}
\begin{enumerate}
\item Initial imputation:
\begin{enumerate}[(a)]
\item[(a)] Impute missing values with RCMimpute assuming  $\Sig = \mathbf{I}$.
\item[(b)] Impute missing values with RCMimpute assuming $\Delt = \mathbf{I}$.
\item[(c)] Average the two estimates.
\end{enumerate}
\item Find the MLE's of the transposable regularized covariance model,
$\hat{\nu}$, $\hat{\mu}$, $\hat{\Sig}$ and
$\hat{\Delt}$ with the imputed missing values fixed.
\item Set the missing values to their conditional expectation given
these parameters:
$\hat{\X}_{m} = \E  ( \X_{m} | \X_{o}, \hat{\nu}, \hat{\mu},
\hat{\Sig}, \hat{\Delt}  )$.
\end{enumerate}
\end{algorithm}

Before discussing the calculations necessary in the final step of the
algorithm, we pause to note a major advantage of our one-step
method.  If the sets of missing values from the marginal models using
RCMimpute are
saved in the first step, then TRCMAimpute can give three sets of
missing value estimates.  Since it is often unknown whether a given
data set may have independent rows or columns, cross-validation, for
example, can be used to determine whether penalizing the covariances
of the rows, columns or both is best for missing value imputation.
This is discussed in detail in the~\hyperref[suppA]{Supplementary Materials}.

%s4.2.2 ###
\subsubsection{Conditional expectations}

We now discuss the final conditional expectation step of our
one-step approximation algorithm.
Recall that the conditional expectation can be computed via
\eqref{full_mn_cond}, but this requires inverting $\Omeg$ and is
therefore avoided.  Instead, we exploit a property of the
mean-restricted matrix-variate normal, namely, that all
marginals of our model are multivariate normal.  This allows us to
find the conditional distributions in a two step process given by
Theorem~\ref{mn_conds}.
%t2
\begin{theorem}\label{mn_conds}
Let $\X \sim N_{n,p}  (\nu, \mu, \Sig, \Delt  )$, $\M = \nu
\mathbf{1}^{T} + \mu^{T} \mathbf{1}$ and partition $\X$, $\M$, $\Sig$,
$\Delt$ as
\begin{eqnarray*} \X &=&  \pmatrix{
\X_{i,m_{i}} & \X_{i,o_{i}} \vspace*{2pt}\cr
\X_{k,m_{i}} & \X_{k,o_{i}}
} =  \pmatrix{
\X_{m_{j},j} & \X_{m_{j},l} \vspace*{2pt}\cr
\X_{o_{j},j} &  \X_{o_{j},l}
},\qquad
\M =  \pmatrix{
\M_{i, r} \vspace*{2pt}\cr
\M_{k, r}
} =  \pmatrix{ \M_{c, j} & \M_{c, l}
},  \\
\Sig &=&  \pmatrix{
\Sig_{i,i} & \Sig_{i, k} \vspace*{2pt}\cr
\Sig_{k, i} & \Sig_{k,k} }, \quad\mbox{and}\quad
\Delt =  \pmatrix{
\Delt_{j,j} & \Delt_{j, l} \vspace*{2pt}\cr
\Delt_{l,j} & \Delt_{l,l } },
\end{eqnarray*}
where $i$ and $j$ denote indices of a row and column, respectively,
$k$ and $l$ are vectors of indices of length $n-1$ and $p-1$,
respectively, and $m_{i}$ and $o_{i}$ denote vectors of indices within
row $i$ and $m_{j}$ and $o_{j}$ indices within column $j$.
\\
Define
\begin{eqnarray*}
\psi &=& \M_{i, r} + \Sig_{i, k} \Sig^{-1}_{k,k}  (\X_{k,r} - \M_{k, r}  ),\qquad
\eta = \M_{ c, j} +  ( \X_{c, l} - \M_{c, l}  ) \Delt^{-1}_{l, l} \Delt_{l,j},  \\
\bolds{\Gamma} &=&  [ \Sig_{i,i} - \Sig_{i, k}
\Sig^{-1}_{k, k} \Sig_{k, i}  ] \otimes \Delt,  \quad\mbox{and}\quad
\bolds{\Phi} = \Sig \otimes  [ \Delt_{j,j} - \Delt_{j, l}
\Delti_{l,l} \Delt_{l, j}  ].
\end{eqnarray*}
Partition $\psi$, $\eta$, $\bolds{\Gamma}$ and $\bolds{\Phi}$ as
$\psi =  { \psi_{m_{i}}\choose \psi_{o_{i}} },
\eta =  (  \eta_{m_{j}}\  \eta_{o_{j}}  )$, \\
\begin{eqnarray*}
\bolds{\Gamma} =  \pmatrix{
\bolds{\Gamma}_{m_{i},m_{i}} & \bolds{\Gamma}_{m_{i},o_{i}} \cr
\bolds{\Gamma}_{o_{i},m_{i}} & \bolds{\Gamma}_{o_{i},o_{i}}
},  \quad\mbox{and}\quad
\bolds{\Phi} =  \pmatrix{
\bolds{\Phi}_{m_{j},m_{j}} & \bolds{\Phi}_{m_{j},o_{j}} \cr
\bolds{\Phi}_{o_{j},m_{j}} & \bolds{\Phi}_{o_{j},o_{j}}
}.
\end{eqnarray*}
Then,
\begin{enumerate}[]
\item[(a)] \mbox{}\vspace*{-21pt}
\begin{eqnarray*}
\hspace*{13pt}&& ( \X_{i,m_{i}} |
\X_{i,o_{i}}, \X_{k, r}  ) \\
&&\qquad\sim N  \bigl( \psi_{m_{i}} +
\bolds{\Gamma}_{m_{i},o_{i}} \bolds{\Gamma}^{-1}_{o_{i},o_{i}}
( \X_{i,o_{i}} - \psi_{o_{i}}  ),
\bolds{\Gamma}_{m_{i},m_{i}} -
\bolds{\Gamma}_{m_{i},o_{i}} \bolds{\Gamma}^{-1}_{o_{i},o_{i}}
\bolds{\Gamma}_{o_{i},m_{i}}   \bigr).
\end{eqnarray*}
\item[(b)]\mbox{}\vspace*{-19pt}
\begin{eqnarray*}
\hspace*{13pt}&& ( \X_{m_{j},j} |\X_{o_{j,j}}, \X_{c,l}  ) \\
&& \qquad
\sim N  \bigl( \eta_{m_{j}} +\bolds{\Phi}_{m_{j},o_{j}} \bolds{\Phi}^{-1}_{o_{j},o_{j}}
( \X_{o_{j},j} - \eta_{o_{j}}  ),
\\
&&\qquad\hspace*{40pt}
\bolds{\Phi}_{m_{j},m_{j}} -
\bolds{\Phi}_{m_{j},o_{j}} \bolds{\Phi}^{-1}_{o_{j},o_{j}}
\bolds{\Phi}_{o_{j},m_{j}}   \bigr).\hspace*{116pt}
\end{eqnarray*}
\end{enumerate}
\end{theorem}
\begin{pf}
See~\hyperref[suppA]{Supplementary Materials}.
\end{pf}

Thus, from Theorem~\ref{mn_conds}, the conditional distribution of
values in a row or column given the rest of the matrix can be
calculated in a two step process where each step takes at most the
number of computations as required for calculating multivariate conditional
distributions.  The first step finds the distribution
of an entire row or column conditional on the rest of the matrix, and the second step finds the conditional
distribution of the values of interest within the row or column.  By
splitting the calculations in this manner, we avoid inverting the $np
\times np$ Kronecker product covariance.  This alternative form for
the conditional distributions of elements in a row or column leads to
an iterative algorithm for calculating the conditional expectation of
the missing values given the observed values.   We call this the Alternating
Conditional Expectations
Algorithm, given in Algorithm~\ref{mn_expec}.

\begin{algorithm}[b]%[!h]
\caption{Alternating Conditional Expectations Algorithm}\label{mn_expec}
\begin{enumerate}
\item Initialize $\hat{\X}^{(0)}_{i,j} = \hat{\nu}_{i} +
\hat{\mu}_{j}$ for $\X_{i,j} \in \X_{m}$.
\item For each row, $i$, with missing values:
\begin{itemize}
\item Set $\hat{\X}^{(k+1)}_{i,m_{i}}  = \E  (
\X_{i,m_{i}} | \X^{(k)}_{i,o_{i}},
\X^{(k)}_{\neq i, r}  )$.
\end{itemize}
\item For each column, $j$, with missing values:
\begin{itemize}
\item Set $\hat{\X}^{(k+1)}_{m_{j},j}  = \E  ( \X_{m_{j},j} |
\X^{(k)}_{o_{j},j}, \X^{(k)}_{c, \neq j}  )$.
\end{itemize}
\item Repeat Steps 1 and 2 until convergence.
\end{enumerate}
\end{algorithm}

%t3
\begin{theorem}\label{e_approx}
Let $\X \sim N_{n,p}  (\nu, \mu, \Sig, \Delt  )$ and
partition
$\operatorname{vec}(\X) =\break  ( \operatorname{vec}(\X_{m})\ \operatorname{vec}(\X_{o})  )$ where $m$ and $o$ are indices
partitioned by rows ($m_{i}$ and $o_{i}$) and columns ($m_{j}$
and $o_{j}$), so that a row $\X_{i, r} =  ( \X_{i,m_{i}}\  \X_{i,o_{i}}  )$ and a column $\X_{c,j} =
{ \X_{m_{j},j} \choose \X_{o_{j},j} }$.  Then,
the Alternating Conditional Expectations Algorithm, Algorithm~\textup{\ref{mn_expec}}, converges to $\E  ( \X_{m} | \X_{o}  )$.
\end{theorem}

\begin{pf}
See~\hyperref[suppA]{Supplementary Materials}.
\end{pf}

Theorem~\ref{e_approx} shows that the conditional expectations needed
in Step 3 of the one-step approximation algorithm can be calculated in
an iterative manner from the conditional distributions of elements in a
row and column, as in Algorithm~\ref{mn_expec}.  Thus, Theorems
\ref{mn_conds} and~\ref{e_approx}
mean that the conditional expectations can be
calculated by separately inverting the row and column covariance
matrices, instead of
the overall Kronecker product covariance.
This reduces the order of
operations from around $O(n^{3} p^{3})$ to $O(n^{3} + p^{3})$, a
substantial savings.  In addition, if both the covariance estimates
and their inverses are known, then one can use the properties of the
Schur complement to further speed computation.  For extremely sparse
matrices or data with few missing elements, the order of operations is
nearly linear in $n$ and $p$ (See~\hyperref[suppA]{Supplementary Materials}).
We also note that the structure of the Alternating
Conditional Expectations Algorithm often leads to a faster rate of convergence,
as discussed in the proof of Theorem~\ref{e_approx}.
For high-dimensional data, these two results mean that
matrix-variate models can be used in any application where
multivariate models are computationally feasible, thus opening the
door to applications of transposable models!

%s4.2.3 ###
\subsubsection{Numerical comparisons}\label{section_approx_res}

\begin{figure}

\includegraphics{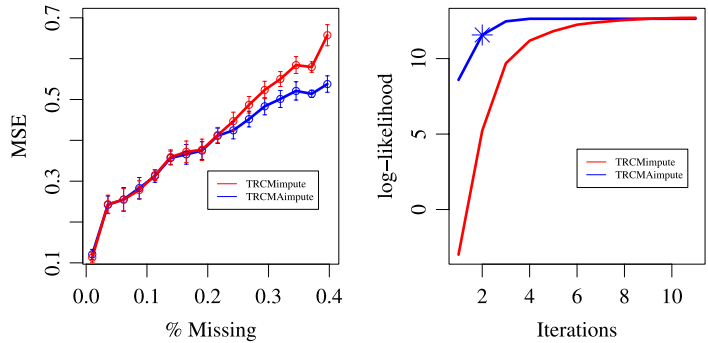}

\caption{Comparison of the mean MSE with standard errors (left)
of the MCECM imputation
algorithm (\mbox{TRCMimpute}) and the one-step
approximation (TRCMAimpute) for transposable data of dimension
$25 \times 25$ with various
percentages of missing data.  Fifty data sets were simulated from
the matrix-variate normal distribution
with autoregressive covariances as given in Section~\protect\ref{section_approx_res}.
Observed log-likelihood (right) verses
iterations for \mbox{TRCMimpute} and TRCMAimpute with 25\% missing
values.  The one-step approximation begins at the TRCM parameter
estimates
using the imputed values from RCMimpute.  The observed log-likelihood
of one-step approximation is
given by a star after the first step.  All methods use $L_{2}$
penalties with $\rho_{r} = \rho_{c} = 1$ for comparison purposes.}\label{fig_approx}
\end{figure}

We now investigate the accuracy of the one-step approximation
algorithm in terms of observed log-likelihood and imputation accuracy
with a numerical example.  Here, we simulate fifty data sets, $25
\times 25$, from the matrix-variate normal model with autoregressive
covariance matrices:
\begin{itemize}
\item Autoregressive: $\Sig_{ij} = 0.8^{|i-j|}$ and $\Delt_{ij} =
0.6^{|i-j|}$.
\end{itemize}
We delete values at random according to certain percentages and report
the mean MSE for both the MCECM algorithm, \mbox{TRCMimpute}, and the
one-step approximation, TRCMAimpute on the right in Figure~\ref{fig_approx}.  The one-step approximation performs comparably, or
slightly better, in
terms of imputation error to the MCECM algorithm for all percentages
of missing values.  We note that TRCMAimpute could give better missing
value estimates if the MCECM algorithm converges to a sub-optimal
stationary point of the observed log-likelihood.
For a data set with 25\% missing values, we
apply the MCECM algorithm and also apply our approximation extended beyond the
first step, but denote the observed log-likelihood after the first step
with a star on the right in Figure~\ref{fig_approx}.  This shows that using
marginals to provide a good starting value does indeed start the
algorithm at a higher observed log-likelihood.  Also, after the first
step, the observed log-likelihood is very close to the fully-iterated
maximum.  Thus, the one-step approximation appears to be a comparable
approximation to the \mbox{TRCMimpute} approximation which is feasible for
use with high-dimensional data sets.

%f1 ###

%s5 ###
\section{Results and simulations}\label{section_res_sims}

The following results indicate that imputation with
transposable regularized covariance models is useful in a
variety of situations and data types, often giving much better error rates
than existing methods.  We first assess the performance of our
one-step approximation, TRCMAimpute, under a variety of simulations
with both full and sparse
covariance matrices.  Authors have suggested
that microarrays and user-ratings data, such as the Netflix
movie-rating data, are transposable or matrix-distributed [\citet{brad}; \citet{bellkor}],
hence, we also assess the performance of our  methods on
these types of data sets.  We compare
performances to three commonly used single imputation methods---SVD methods
(SVDimpute),
$k$-nearest neighbors (KNNimpute) and local least squares (LLSimpute)
[\citet{svdknn}; \citet{lls}].  For the SVD method, we use a reduced rank
model with a column mean effect.  The rank of the SVD is determined
by cross-validation; regularization is not used on the singular
vectors so that only one parameter is needed for selection by
cross-validation.  For $k$-nearest neighbors and local least squares
also, a column mean
effect is used and the number of neighbors, $k$, is selected
via cross-validation.  If the number of observed elements is limited,
the pairwise-complete correlation matrix is used to determine the
closest neighbors.

%s5.1 ###
\subsection{Simulations}\label{section_sims}

We test our imputation method for transposable data
under a variety of
simulated distributions, both multivariate and matrix-variate.  All
simulations use one of four covariance types given below.  These are
numbered as they appear in the simulation table:
\begin{enumerate}
\item Autoregressive: $\Sig_{ij} = 0.8^{|i-j|}$ and $\Delt_{ij} =
0.6^{|i-j|}$.
\item Equal off-diagonals: $\Sig_{ij} = 0.5$ and $\Delt_{ij} = 0.5$ for
$i \neq j$, and $\Sig_{ii} = 1$ and \mbox{$\Delt_{ii} = 1$}.
\item Blocked diagonal: $\Sig_{ii} = 1$ and $\Delt_{ii} = 1$ with
off-diagonal elements of $5 \times 5$ blocks
of $\Sig$ are 0.8 and of $\Delt$, 0.6.
\item Banded off-diagonals: $\Sig_{ii} = 1$ and $\Delt_{ii} = 1$ with
%{\small
\begin{eqnarray*}
\Sig_{ij} &=& \cases{
0.8, &\quad \mbox{if }$|i-j|$\mbox{ divisible by 5}, \cr
0, &\quad \mbox{otherwise}.
}
\\
\Delt_{ij} &=& \cases{
0.6, &\quad \mbox{if }$|i-j|$\mbox{ divisible by 5}, \cr
0, &\quad \mbox{otherwise}.
}
\end{eqnarray*}
%}
\end{enumerate}
The first simulation, with results in Table~\ref{tab_sims}, compares
performances with both multivariate distributions, only $\Sig$ given, and
matrix-variate distributions, both $\Sig$ and $\Delt$ given.  In these
simulations, the data is of dimension $50 \times 50$ with either 25\%
or 75\% of the values missing at random.    The simulation
given in Table~\ref{tab_long} gives results for matrix-variate
distributions with one dimension much larger than the other, $100
\times 10$ and 10\% of values missing at random.  The final
simulation, in Table~\ref{tab_robust}, tests the performance of our
method when the data has a transposable covariance structure, but is
not normally distributed.  Here, the data, of dimension $50 \times 50$
with 25\% of values missing at random,
is either distributed Chi-square with three degrees of freedom or
Poisson with mean three.  The Chi-square and Poisson distributions
introduce large outliers and the Poisson distribution is discrete.
All three sets of simulations are compared to SVD imputation and $k$-nearest
neighbor imputation.

%t1 ###
\begin{table}
\caption{Mean MSE with standard error computed over 50 data sets of dimension
$50 \times 50$ simulated under the matrix-variate normal
distribution with  covariances given in
Section~\protect\ref{section_sims}.  In the upper portion of the table, 25\%
of values are missing and in the lower, 75\% missing.  The TRCM one-step approximation with
$L_{1}\dvtx L_{1}$, $L_{1}\dvtx L_{2}$  and $L_{2}\dvtx L_{2}$ penalties was used
as well as the SVD
and $k$-nearest neighbor imputation.  Below the errors for
TRCMAimpute, we give the number of simulations out of 50 in which a marginal,
multivariate method (RCMimpute) was chosen over the matrix-variate method.
Parameters were chosen for
all methods via 5-fold cross-validation.  Best performing methods are
given in bold}\label{tab_sims}
\tabcolsep=0pt
\begin{tabular*}{\textwidth}{@{\extracolsep{\fill}}l lllll@{}}
\hline
& \multicolumn{3}{c}{\textbf{TRCMAimpute}} & \multicolumn{2}{c@{}}{\textbf{Others}} \\
& \multicolumn{3}{c}{\hrulefill} & \multicolumn{2}{c@{}}{\hrulefill} \\
& \multicolumn{1}{c}{$\bolds{L_{1}\dvtx L_{1}}$}
& \multicolumn{1}{c}{$\bolds{L_{1}\dvtx L_{2}}$}
& \multicolumn{1}{c}{$\bolds{L_{2}\dvtx L_{2}}$}
& \multicolumn{1}{c}{\textbf{SVD}}
& \multicolumn{1}{c@{}}{\textbf{KNN}} \\
\hline
$\Sig_{1}$   & 0.8936\ (0.01) & 0.725\ (0.0069) & \textbf{0.5919}\ (0.0056) & 0.634\ (0.0081) & \textbf{0.448}\ (0.005)\\
& \multicolumn{1}{c}{45$/$50} & \multicolumn{1}{c}{0$/$50} & \multicolumn{1}{c}{50$/$50} & &
\\[3pt]
$\Sig_{1}$, $\Delt_{1}$  & 0.8255\ (0.012)& 0.6315\ (0.0078) & \textbf{0.5402}\ (0.0067) & \textbf{0.4603}\ (0.0083)& 0.8034\ (0.016)\\
& \multicolumn{1}{c}{0$/$50} & \multicolumn{1}{c}{0$/$50} & \multicolumn{1}{c}{0$/$50} & &
\\[3pt]
$\Sig_{2}$  & 0.895\ (0.016) & \textbf{0.7829}\ (0.013)& \textbf{0.6392}\ (0.008)& 0.993\ (0.019)& 0.9498\ (0.017)\\
& \multicolumn{1}{c}{43$/$50} & \multicolumn{1}{c}{0$/$50} & \multicolumn{1}{c}{48$/$50} & &
\\[3pt]
$\Sig_{2}$, $\Delt_{2}$ & 0.749\ (0.044)&   0.6867\ (0.034)&\textbf{0.4556}\ (0.0098)&   \textbf{0.6821}\ (0.051)& 0.8273\ (0.055)\\
& \multicolumn{1}{c}{0$/$50}  & \multicolumn{1}{c}{0$/$50}  & \multicolumn{1}{c}{48$/$50} &  &
\\[3pt]
$\Sig_{3}$ & 1.04\ (0.017)  & 1.02\ (0.017)&0.9348\ (0.016) & \textbf{0.7384}\ (0.012)& \textbf{0.9115}\ (0.014)\\
&  \multicolumn{1}{c}{37$/$50} & \multicolumn{1}{c}{9$/$50} & \multicolumn{1}{c}{49$/$50} &    &
\\[3pt]
$\Sig_{3}$, $\Delt_{3}$ & 1.012\ (0.02) & 0.9477\ (0.019)&\textbf{0.8585}\ (0.017)&  \textbf{0.7271}\ (0.016)& 0.9886\ (0.019)\\
& \multicolumn{1}{c}{5$/$50} & \multicolumn{1}{c}{0$/$50}  & \multicolumn{1}{c}{37$/$50} &  &
\\[3pt]
$\Sig_{4}$  & 0.9986\ (0.018) & 0.9407\ (0.017) & \textbf{0.8067}\ (0.014)& \textbf{0.4903}\ (0.0076)& 0.9057\ (0.014)\\
&  \multicolumn{1}{c}{37$/$50} & \multicolumn{1}{c}{3$/$50}  & \multicolumn{1}{c}{48$/$50} &  &
\\[3pt]
$\Sig_{4}$, $\Delt_{4}$ & 0.9726\ (0.033) & 0.855\ (0.028)& \textbf{0.6999}\ (0.022) & \textbf{0.5282}\ (0.024)& 0.9366\ (0.031)\\
&  \multicolumn{1}{c}{6$/$50} & \multicolumn{1}{c}{0$/$50}  &  \multicolumn{1}{c}{39$/$50} &  &
\\[6pt]
$\Sig_{1}$  & 0.9134\ (0.0096) & \textbf{0.9083}\ (0.0092) & \textbf{0.8948}\ (0.009)& 1.173\ (0.013)& 0.9349\ (0.0092)\\
& \multicolumn{1}{c}{21$/$50} &  \multicolumn{1}{c}{18$/$50} &  \multicolumn{1}{c}{21$/$50} &  &
\\[3pt]
$\Sig_{1}$, $\Delt_{1}$  & 0.867\ (0.011) & \textbf{0.8569}\ (0.01)&\textbf{0.845}\ (0.0096)&  0.9535\ (0.01)& 0.9736\ (0.013)\\
&  \multicolumn{1}{c}{0$/$50} & \multicolumn{1}{c}{0$/$50}  & \multicolumn{1}{c}{0$/$50}  &  &
\\[3pt]
$\Sig_{3}$ & 1.053\ (0.01) & 1.052\ (0.01) & \textbf{1.048}\ (0.01)&1.22\ (0.013)& \textbf{1.03}\ (0.01)\\
&  \multicolumn{1}{c}{7$/$50} & \multicolumn{1}{c}{7$/$50}  & \multicolumn{1}{c}{7$/$50}  &  &
\\[3pt]
$\Sig_{3}$, $\Delt_{3}$  & 1.001\ (0.014) & \textbf{0.9968}\ (0.014)&\textbf{0.9945}\ (0.014) & 1.11\ (0.016)& 1.006\ (0.014)\\
& \multicolumn{1}{c}{0$/$50}  & \multicolumn{1}{c}{0$/$50}  & \multicolumn{1}{c}{1$/$50} &  &   \\
\hline
\end{tabular*}
\end{table}

%t2 ###
\begin{table}
\caption{Mean MSE with standard errors over 50 data sets of
dimension $100 \times 10$ with 10\% missing values
simulated under the matrix-variate normal with covariances given in
Section~\protect\ref{section_sims}.  The TRCM one-step approximation with
$L_{2}\dvtx L_{1}$  and $L_{2}\dvtx L_{2}$ penalties was used as well as the SVD
and $k$-nearest neighbor imputation.  Parameters were chosen for
all methods via 5-fold cross-validation.  Best performing methods are
given in bold}\label{tab_long}
\begin{tabular*}{\textwidth}{@{\extracolsep{\fill}}l llll@{}}
\hline
& \multicolumn{2}{c}{\textbf{TRCMAimpute}} & \multicolumn{2}{c@{}}{\textbf{Others}}
\\[-6pt]
& \multicolumn{2}{c}{\hrulefill} & \multicolumn{2}{c@{}}{\hrulefill} \\
& \multicolumn{1}{c}{$\bolds{L_{2}\dvtx L_{1}}$}
& \multicolumn{1}{c}{$\bolds{L_{2}\dvtx L_{2}}$}
& \multicolumn{1}{c}{\textbf{SVD}}
& \multicolumn{1}{c@{}}{\textbf{KNN}} \\
\hline
$\Sig_{1}$, $\Delt_{1}$ & 0.8227 (0.019) & 0.7072 (0.016) & 1.075 (0.024) & \textbf{0.6971} (0.018)
\\
$\Sig_{2}$, $\Delt_{2}$ & 1.019 (0.15) & \textbf{0.9441} (0.13) & 1.306 (0.23) & 1.057 (0.17)
\\
$\Sig_{3}$, $\Delt_{3}$ & 0.9372 (0.047) & \textbf{0.841} (0.042) & 1.121 (0.05) & 0.9241 (0.042)
\\
$\Sig_{4}$, $\Delt_{4}$ & 0.7044 (0.059) & \textbf{0.6148} (0.049) & 0.9751 (0.074) & 1.118 (0.089) \\
\hline
\end{tabular*}
\end{table}

%t3 ###
\begin{table}[b]
\caption{Mean MSE with standard error computed over 50 data sets of dimension
$50 \times 50$ with 25\% missing values simulated under the Chi-square distribution with 3
degrees of freedom or the Poisson distribution with mean 3 with
Kronecker product covariance structure given by the covariances in
Section~\protect\ref{section_sims}.  The TRCM one-step approximation with
$L_{2}\dvtx L_{1}$  and $L_{2}\dvtx L_{2}$ penalties was used as well as the SVD
and $k$-nearest neighbor imputation.  Parameters were chosen for
all methods via 5-fold cross-validation.  Best performing methods are
given in bold}\label{tab_robust}
\begin{tabular*}{\textwidth}{@{\extracolsep{\fill}}lc llll@{}}
\hline
& & \multicolumn{2}{c}{\textbf{TRCMAimpute}} & \multicolumn{2}{c@{}}{\textbf{Others}}
\\[-6pt]
& & \multicolumn{2}{c}{\hrulefill} & \multicolumn{2}{c@{}}{\hrulefill} \\
& & \multicolumn{1}{c}{$\bolds{L_{2}\dvtx L_{1}}$}
& \multicolumn{1}{c}{$\bolds{L_{2}\dvtx L_{2}}$}
& \multicolumn{1}{c}{\textbf{KNN}}
& \multicolumn{1}{c@{}}{\textbf{SVD}} \\
\hline
Chi-square & $\Sig_{1}$, $\Delt_{1}$ &  3.824 (0.065) & \textbf{2.611} (0.044) & \phantom{0}6.85 (0.34) & \phantom{0}7.684 (0.15) \\
& $\Sig_{3}$, $\Delt_{3}$ &  5.525 (0.14) & \textbf{5.068} (0.15) & 29.41 (0.83) & 50.16 (0.74)
\\[3pt]
Poisson  & $\Sig_{1}$, $\Delt_{1}$ &  2.442 (0.05) & \textbf{1.571} (0.021) &\phantom{0}8.04 (0.34) & \phantom{0}5.824 (0.11) \\
& $\Sig_{3}$, $\Delt_{3}$   & 3.045 (0.075) & \textbf{2.813} (0.081) &29.13 (0.95) & 49.2 (0.68) \\
\hline
% $\Sig_{1}$, $\Delt_{1}$ & Chi-square & 1.53 (0.028) & 1.026 (0.014) &
% 1.519 (0.03) & 2.305 (0.045) \\
% \hline
% $\Sig_{3}$, $\Delt_{3}$ & Chi-square & 0.7592 (0.011) & 0.6859 (0.011)
% & 1.156 (0.017) & 10.94 (0.2) \\
% \hline
% $\Sig_{1}$, $\Delt_{1}$ & Poisson & 0.9744 (0.018) & 0.6659 (0.01) &
% 0.9235 (0.017) & 1.322 (0.028) \\
% \hline
% $\Sig_{3}$, $\Delt_{3}$ & Poisson & 0.4703 (0.0069) & 0.4333 (0.0076)
% & 0.7778 (0.014) & 11.23 (0.19) \\
% \hline
\end{tabular*}
\end{table}

These simulations show that TRCMAimpute is
competitive with two of the most commonly used single imputation
methods, SVD and $k$-nearest neighbor imputation.  First, TRCM with
$L_{2}$ penalties outperforms the other possible TRCM penalty types.
This may be due to the fact that the covariance estimates with $L_{2}$
penalties has a globally unique solution, Theorem~\ref{l2l2_sol},
while the estimation procedure for other penalty types only reaches a
stationary point, Proposition~\ref{cw_conv}. The one-step approximation permits
the flexibility to choose either multivariate or transposable
models.   As seen with
smaller percentages of missing values, cross-validation generally
chooses the correct model for the $L_{1}\dvtx L_{1}$ penalty-type, but
seems to prefer the marginal multivariate models for the $L_{2}\dvtx L_{2}$
penalties.  However, with 75\% of the values missing,
the transposable model is often chosen even if the underlying distribution
is multivariate.  The additional structure of the TRCM covariances may
allow for more information to be gleaned from the few observed values,
perhaps explaining the better performance of the matrix-variate
model.  TRCMAimpute seems to perform best in comparison to SVD and
$k$-nearest neighbor imputation for the full covariances with equal
off-diagonal elements.  Our TRCM-based imputation methods appear
particularly robust to departures from normality and perform well even
in the presence of large
outliers, as shown in Table
\ref{tab_robust}.
Overall, imputation
methods based on transposable covariance models compare favorably in
these simulations.

%s5.2 ###
\subsection{Microarray data}\label{section_micro}

Microarrays are high-dimensional matrix-data that often contain
missing values.  Usually, one assumes that the genes are correlated
while the arrays are independent.  Efron questions this assumption,
however, and suggests using a matrix-variate normal model
[\citet{brad1}].  Indeed, the matrix-variate framework, and, more
specifically, the TRCM model seem appropriate models for microarray
data for several reasons.  First, one usually centers both the genes
and the arrays before analysis, a structure which is built in to our
model.  Second, TRCMs have the ability to span many models which
include a marginal model where the rows are distributed as a
multivariate normal and the arrays are independent.  Hence,
if a microarray is truly multivariate, our model can accommodate
this. But, if there are true correlations within the arrays, TRCM can
appropriately measure this correlation and account for it when
imputing missing values.  Last, the graphical nature of our model can estimate
the gene network and then use this information to more accurately
estimate missing data.

For our analysis, we use a microarray data set of kidney
cancer tumor samples [\citet{kidney}].
The data set contains 14,814 genes and 178 samples.  About 10\% of the
data is missing.  For the following figures, all of
the genes with no missing values were taken, totaling 1031 genes.
Missing values were then placed at random.  Errors were assessed by
comparing the
imputed values to the true observed values.

%f2 ###
\begin{figure}

\includegraphics{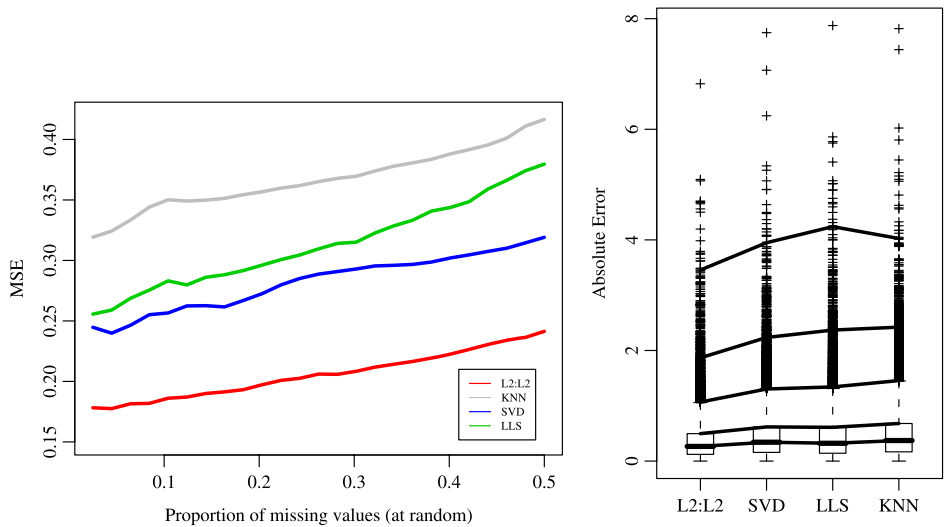}

\caption{Left: Comparison of MSE for imputation methods on kidney
cancer microarray data with
different proportions of missing values.  Genes in which all
samples are observed are taken with values deleted at
random.  TRCMAimpute, $L_{2}\dvtx L_{2}$ and common imputation
methods KNNimpute, SVDimpute and LLSimpute are compared with
all parameters chosen by 5-fold cross-validation.
Cross-validation chose to penalize only the arrays for the
one-step approximation algorithm, \mbox{TRCMAimpute}. Right: Boxplots
of individual absolute errors for various
imputation methods.  Genes in which all
samples are observed were taken and deleted in the same pattern
as a random gene in the original data set.  Lines
are drawn at the 50\%, 75\%, 95\%, 99\% and 99.9\% quantiles.
\mbox{TRCMAimpute}, $L_{2}\dvtx L_{2}$ has a mean absolute
error of 0.37 and has lower errors at every quantile than
its closest competitor, SVDimpute, which has a mean absolute
error of 0.46.
}\label{fig_micro_pmiss}
\end{figure}

We assess the performance of TRCM imputation methods on this microarray
data and compare them to existing methods for various percentages of
missing values, deleted at random, on the right in Figure
\ref{fig_micro_pmiss}.  Here, we use $L_{2}$
penalties since these are computationally less expensive for
high-dimensional data.  \mbox{TRCMAimpute} outperforms competing methods in
terms of imputation error for all percentages of missing values.  We
note that cross-validation exclusively chose the marginal,
multivariate model from the one-step approximation.  This indicates
that the arrays in this microarray data set may indeed by independent.

Often, microarray data sets are not missing randomly.  Also,
researchers are interested in not only the error in
terms of MSE, but the individual errors made as well.  To investigate
these issues, we assess individual absolute errors of data that is
missing in the same pattern as the original data.  For each
complete gene, values were set to missing in the
same arrays as a randomly sampled gene from the original data set.
The right panel of Figure
\ref{fig_micro_pmiss} displays the boxplots of the absolute imputation errors.
Lines are drawn at quantiles to assess the relative performances
of each method.  Here, TRCMAimpute has
lower absolute errors at each quantile.  Also, the set of imputed
values has far
fewer outliers than competing methods.  The mean absolute error for
TRCMAimpute is 0.37, far below the next two methods, LLSimpute and
SVDimpute which have a mean absolute error of 0.46.
Altogether, our results illustrate the utility and flexibility of
using TRCMs for
missing value imputation in microarray data.

%s5.3 ###
\subsection{Netflix data}\label{section_netflix}

We compare transposable
regularized covariance models and existing methods on the Netflix movie
rating data [\citet{netflix}].  The TRCM framework seems well-suited to
model this user-ratings data.  As discussed in the \hyperref[section_intro]{Introduction}, our
model allows for not only correlations among both the customers and
movies, but also between them as well.  In addition, TRCM models the graph
structure of the customers and the movies.  Thus, we can fill in a
customer's rating of a particular movie based on the customer's links
with other customers and the movie's links with other movies.  Also,
many have noted that the unrated movies in the Netflix data are not
simply missing at random and may contain meaningful information.
A customer, for example, may not have rated a movie because the movie
was not of interest and, thus, they never saw it.  While it may appear
that our method requires a missing at random assumption, this is not
necessarily the case.  When two customers have similar sets of unrated
movies, after removing the means, our algorithm begins with the
unrated movies set to zero.  Thus, these two customers would exhibit
high correlation simply due to the pattern of missing values.  This
correlation could yield an estimated ``link'' between the customers in
the inverse covariance matrix.  This would then be used to estimate
the missing ratings.  Hence, our method
can find relationships between sets of missing values and use these to
impute the missing values.

\begin{figure}

\includegraphics{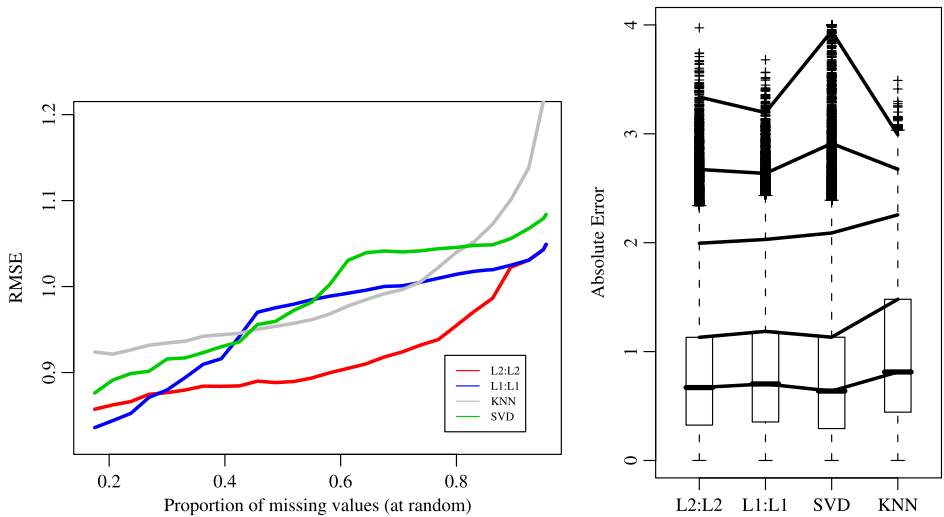}

\caption{Left: Comparison of the root MSE (RMSE) for a subset of the
Netflix data for \mbox{TRCMAimpute}, $L_{2}\dvtx L_{2}$ and $L_{1}\dvtx L_{1}$,
to KNNimpute and SVDimpute.  A dense subset
was obtained by ranking the movies and
customers in terms of number of ratings and taking the top 250
movies and 250 customers.  This subset has around 12\% missing
and additional values were deleted at random, up to 95\%.  With
95\% missing, the RMSE of TRCMAimpute is 1.049 compared to 1.084
of the SVD and 1.354 using the movie averages.
Right: Boxplots of
absolute errors for the dense subset with missing entries in the
pattern of the original data.  Customers with at least one ranking
out of the 250 movies were selected at random and entries were
deleted according to these customers leaving 74\% missing.
Quantiles of the absolute errors are shown at 50\%, 75\%, 95\%,
99\% and 99.9\%.  The RMSE of the methods are as follows
$L_{2}\dvtx L_{2}$: 1.005, $L_{1}\dvtx L_{1}$: 1.029, SVD: 1.032, KNN: 1.184. }\label{fig_netflix}
\end{figure}

The Netflix data set is extremely high-dimensional, with
over 480,000 customers and over 17,000 movies, and is
very sparse, with over 98\% of the ratings missing.  Hence, assessing
the utility of our methods from this data as a whole is not currently feasible.
Instead, we rank both the movies and the customers by the number of
ratings and take as a subset the top 250 customer's ratings of the
top 250 movies.  This subset has around 12\% of the ratings missing.
We then delete more data at random to evaluate the performance of the
methods.  In addition, for each customer in this subset ratings were
deleted for movies corresponding to the unrated movies of a randomly
selected customer with at least one rating out of the 250 movies.
This leaves 74\% of ratings missing.
Figure~\ref{fig_netflix} compares the performances of the
TRCM methods to existing methods for both subsets with both missing at
random and missing in the pattern of the original values.

%f3 ###

Before discussing these results, we first make a note about the
comparability of our errors rates to those for the Netflix Prize
[\citet{netflix}].  Because we
chose the subset of data based on the number of observed
ratings, we can expect the RMSE to be higher here than applying these
methods to the full data set.  This method of obtaining a subset leaves
out potentially thousands of highly correlated customers or movies
that would greatly increase a method's predictive ability.  In fact,
the RMSE of the SVD method on the entire Netflix data is 0.91
[\citet{netflixrbm}], much less than the observed RMSE of 1.084 for the
SVD on our
subset with 95\% missing.  Thus, we can conjecture that all of the methods we present
would do better in terms of RMSE using the entire data set than the
small subset on which we present results.

The results indicate that TRCM imputation methods, particularly with
$L_{2}$ penalties, are competitive
with existing methods on the missing at random data.  At higher percentages of missing values, our
methods perform notably well.
With 95\% missing values in our subset, TRCMAimpute has a RMSE of 1.049
compared to the SVD at 1.084 and 1.354 using the movie averages.   This
is of potentially great interest for missing data imputation on a
larger scale where the percentage of missing data is greater than
98\%.  We note that at smaller percentages of missing values, marginal
models penalizing the movies were often chosen by cross-validation,
indicating that the movies may have more predictive power, whereas,
at larger percentages of missing values, cross-validation chose to
penalize both the rows and the columns, indicating that possibly more
information can be gleaned from few observed values using transposable
methods.

Our methods also preform well when the data is missing in the same
pattern as the original.  The $L_{2}:L_{2}$ method had the best
results with a RMSE of 1.005 followed by $L_{1}:L_{1}$ with 1.029, SVD
with 1.032 and $k$-nearest neighbors with 1.184.  From the boxplots of
absolute values Figure~\ref{fig_netflix} (right), we see that the SVD
has many large outliers in absolute value, while the $L_{1}$ penalties
led to the fewest number of large errors.
Since leading imputation methods for the Netflix Prize are
ensembles of many different methods [\citet{bellkor}], we do not believe
that TRCM methods alone would outperform ensemble methods.  If,
however, our methods outperform other individual methods, they could
prove to be beneficial additions to imputation ensembles.

%s6 ###
\section{Discussion}\label{section_discussion}

We have formulated a parametric model for
matrix-data along with computational advances that allow this model to
be applied to missing value estimation in high-dimensional data sets with possibly
complex correlations between and among the rows and columns.

Our MCECM and one-step approximation imputation approaches are
restricted to data sets where for each pair
of rows, there is at least one column in which both entries are
observed and vice versa for each pair of columns.
A major drawback of TRCM imputation methods is
computational cost.  First, RCMimpute using the columns as features
costs $O(p^{3})$.   This is
roughly on the order of other common imputation methods such as the
SVDimpute which costs $O(np^{2})$.  Our one-step approximation,
TRCMAimpute, using the computations for the Alternating Conditional
Expectations algorithm given in the~\hyperref[suppA]{Supplementary Materials}, costs $O(\sum_{i=1}^{n} \min\{ |m_{i}|,
|o_{i}| \}^{3} + \sum_{j=1}^{p} \min \{ |m_{j}|, |o_{j}|
\}^{3} + n^{3} + p^{3}  )$, where\vspace*{2pt} $|m_{i}|$ and $|o_{i}|$ are the
number of missing and observed elements of row $i$, respectively.

The main application of this paper has been to missing value
imputation.  We note that this is separate from the matrix-completion
methods via convex optimization of Candes and Recht
[\citet{candesmatcomp}], which focuses on matrix-reconstruction instead
of imputation.  Also, we have presented a single imputation procedure,
but our techniques can easily be extended to incorporate multiple
imputation.  We present a repeated imputations approach by taking samples from
the posterior distribution [\citet{rubinmultimp}] with the Bayesian one-step
approximation in the~\hyperref[suppA]{Supplementary Materials}.
In addition, we have not discussed
ultimate use or analysis of the imputed data, which will often dictate
the imputation approach.  Our imputation methods form a
foundation that can be extended to further address these issues.

We also pause to address the appropriateness of the Kronecker product
covariance matrix to model the covariances observed in real data.
While we do not assume that this particular structure is suitable for
all data, we feel comfortable using the model because of its
flexibility.  Recall that all marginal distributions of the
mean-restricted matrix-variate normal are multivariate normal.  This
includes the distribution of elements within a row or column, or the
distribution of elements from different rows or columns.  All of the
marginals of a set of elements are given by the mean and covariance
parameters of the elements' rows and columns.
Thus, our model says that the location of elements within a
matrix determine their distribution, often a reasonable assumption.
Also, if either the covariance
matrix of the rows or the columns is the identity matrix, then we are
back to the familiar multivariate normal model.  This flexibility to
fit numerous multivariate models and to adapt to structure within a
matrix is an important advantage of our matrix-variate model.

Transposable regularized covariance models may be of potential
mathematical and practical interest in numerous fields.  TRCMs
allow for nonsingular estimation of the covariances of the rows and
columns, which is essential for any application.  Adding restrictions
to the mean of the TRCM allows one to estimate all parameters from a
single observed data matrix.  Also, introduction of efficient methods
of calculating conditional distributions and expectations make this
model computationally feasible for many applications.  Hence,
transposable regularized covariance models  have many potential future
uses in areas such as hypothesis testing,
classification and prediction, and data mining.

\section*{Acknowledgments}
Thanks to Steven Boyd for a discussion about the minimization of
biconvex functions.  We also thank two referees and the Editor for
their helpful comments that led to several improvements
in this paper.

\begin{supplement}[id=suppA]
\stitle{Additional methods and proofs}\label{suppA}
\slink[doi]{10.1214/09-AOAS314SUPP}
\slink[url]{http://lib.stat.cmu.edu/aoas/314/supplement.PDF}
\sdatatype{.pdf}
\sdescription{This includes sections on the multivariate imputation
method \textit{RCMimpute}, numerical results on TRCM covariance
estimation, a discussion of properties of the MCECM algorithm for
imputation, computations for the Alternating Conditional
Expectations Algorithm,
a Bayesian one-step approximation to \textit{\mbox{TRCMimpute}} along with a
Gibbs sampling algorithm, discussion of
cross-validation for estimating penalty parameters, and proofs of
theorems and propositions.}
\end{supplement}

\printaddresses

\end{document}